# Finding Botnets Using Minimal Graph Clusterings

**Peter Haider** HAIDER@CS.UNI-POTSDAM.DE
**Tobias Scheffer** SCHEFFER@CS.UNI-POTSDAM.DE
University of Potsdam, Department of Computer Science, August-Bebel-Str. 89, 14482 Potsdam, Germany

## Abstract

We study the problem of identifying botnets and the IP addresses which they comprise, based on the observation of a fraction of the global email spam traffic. Observed mailing campaigns constitute evidence for joint botnet membership, they are represented by cliques in the graph of all messages. No evidence against an association of nodes is ever available. We reduce the problem of identifying botnets to a problem of finding a minimal clustering of the graph of messages. We directly model the distribution of clusterings given the input graph; this avoids potential errors caused by distributional assumptions of a generative model. We report on a case study in which we evaluate the model by its ability to predict the spam campaign that a given IP address is going to participate in.

## 1. Introduction

We address the problem of identifying botnets that are capable of exploiting the internet in a coordinated, distributed, and harmful manner. Botnets consist of computers that have been infected with a software virus which allows them to be controlled remotely by a botnet operator. Botnets are used primarily to disseminate email spam, to stage distributed denial-of-service (DDoS) attacks, and to harvest personal information from the users of infected computers (Stern, 2008).

Providers of computing, storage, and communication services on the internet, law enforcement and prosecution are interested in identifying and tracking these threats. An accurate model of the set of IP addresses over which each existing botnet extends would make it possible to protect services against distributed denial-of-service attacks by selectively denying service requests from the nodes of the offending botnet.

Evaluating botnet models is difficult, because the ground truth about the sets of IP addresses that constitute each botnet at any given time is entirely unavailable (Dittrich & Dietrich, 2008). Many studies on botnet identification conclude with an enumeration of the conjectured number and size of botnets (Zhuang et al., 2008). Reliable estimates of the current size of one particular botnet require an in-depth analysis of the communication protocol used by the network. For instance, the size of the Storm botnet has been assessed by issuing commands that require all active nodes to respond (Holz et al., 2008). However, once the communication protocol of a botnet is understood, the botnet is usually taken down by law enforcement, and one is again ignorant of the remaining botnets.

We develop an evaluation protocol that is guided by the basic scientific principle that a model has to be able to predict future observable events. We focus on email spam campaigns which can easily be observed by monitoring the stream of messages that reach an email service provider. Our evaluation metric quantifies the model's ability to predict which email spam campaign a given IP address is going to participate in.

Previous studies have employed clustering heuristics to aggregate IP addresses that participated in joint campaigns into conjectured botnets (Xie et al., 2008; Zhuang et al., 2008). Because an IP address can be a part of multiple botnets during an observation interval, this approach is intrinsicaly inaccurate. The problem is furthermore complicated as it is possible that a botnet processes multiple campaigns simultaneously, and multiple botnets may be employed for large campaigns. We possess very little background knowledge about whether multiple networks, each of which has been observed to act in a coordinated way, really form one bigger, joint network. Also, distributional assumptions about the generation of the observable events are very hard to motivate. We address this lack of prior knowledge by directly modeling the conditional distri-bution of clusterings given the observable data, and by searching for minimal clusterings that refrain from merging networks as long as empirical evidence does not render joint membership in a botnet likely.

Other studies have leveraged different types of data in order to identify botnets. For example Mori et al. (2010); DiBenedetto et al. (2010) record and cluster fingerprints of the spam-sending hosts' TCP behavior, exploiting that most bot types use their own protocol stacks with unique characteristics. Yu et al. (2010) identify bot-generated search traffic from query and click logs of a search engine by detecting shifts in the query and click distributions compared to a background model. Another angle to detect bots is to monitor traffic from a set of potentially infected hosts and find clusters in their outgoing and incoming packets (Gu et al., 2008; John et al., 2009); for example, DNS requests of bots used to connect to control servers (Choi et al., 2009) or IRC channel activity (Goebel & Holz, 2007). The major difference here is that access to *all* the traffic of the hosts is required, and thus these methods only work for finding infected hosts in a network under one's control.

The rest of this paper is organized as follows. In Section 2, we discuss our approach to evaluating botnet models by predicting participation in spamming campaigns. In Section 3, we establish the problem of minimal graph clustering, devise a probabilistic model of the conditional distribution of clusterings given the input graph, and derive a Gibbs sampler. Section 4 presents a case study that we carried out with an email service provider. Section 5 concludes.

## 2. Problem Setting and Evaluation

The ground truth about the sets of IP addresses that constitute each botnet is unavailable. Instead, we focus on the botnet model's ability to predict observable events. We consider email spam campaigns which are one of the main activities that botnets are designed for, and which we can easily observe by monitoring the stream of emails that reach an email service provider. Most spam emails are based on a campaign template which is instantiated at random by the nodes of a botnet. Clustering tools can identify sets of messages that are based on the same campaign template with a low rate of errors (Haider & Scheffer, 2009). A single campaign can be disseminated from the nodes of a single botnet, but it is also possible that a botnet processes multiple campaigns simultaneously, and multiple botnets may be employed for large campaigns.

We formalize this setting as follows. Over a fixed period of time, $n$ messages are observed. An adjacency matrix $X$ of the graph of messages reflects evidence that pairs of messages originate from the same botnet. An edge between nodes $i$ and $j$—represented by an entry of $X_{ij} = 1$—is present if messages $i$ and $j$ have been sent from the same IP address within the fixed time slice, or if a campaign detection tool has assigned the messages to the same campaign cluster. Both types of evidence are uncertain, because IP addresses may have been reassigned within the time slice and the campaign detection tool may incur errors. The absence of an edge is only very weak and unreliable evidence against joint botnet membership, because the chance of not observing a link between nodes that are really part of the same botnet is strongly dependent on the observation process.

The main part of this paper will address the problem of inferring a reflexive, symmetric *edge selector matrix* $Y$ in which entries of $Y_{ij} = 1$ indicate that the messages represented by nodes $i$ and $j$ originate from the same botnet. The transitive closure $Y^+$ of matrix $Y$ defines a clustering $C_Y$ of the nodes. The clustering places each set of nodes that are connected to one another by the transitive closure $Y^+$ in one cluster; the clustering is the union of clusters:

$$C_Y = \bigcup_{i=1}^{n} \{\{j : Y^+_{ij} = +1\}\}. \quad (1)$$

Because $Y^+$ is reflexive, symmetric and transitive, it partitions all nodes into disjoint clusters; that is, $c \cap c' = \emptyset$ for all $c, c' \in C_Y$, and $\bigcup_{c \in C_Y} c = \{1, \ldots, n\}$.

An unknown process generates future messages which are characterized by two observable and one latent variable. Let the multinomial random variables $s$ indicate the campaign cluster of a newly received message, $a$ indicate the IP address, and let latent variable $c$ indicate the originating cluster, associated with a botnet. We quantify the ability of a model $C_Y$ to predict the observable variable $s$ of a message given $a$ in terms of the likelihood

$$P(s|a, C_Y) = \sum_c P(s|c, C_Y) P(c|a, C_Y). \quad (2)$$

Equation 2 assumes that the distribution over campaigns is conditionally independent of the IP address given the botnet; that is, botnet membership alone determines the distribution over campaigns.

Multinomial distribution $P(s|c, C_Y)$ quantifies the likelihood of campaign $s$ within the botnet $c$. It can be estimated easily on training data because model $C_Y$ fixes the botnet membership of each message. Multinomial distribution $P(c|a, C_Y)$ quantifies the probability that IP address $a$ is part of botnet $c$ given model $C_Y$. Model $C_Y$ assigns each node—that is, message—to a botnet. However, an address can be observed multiple times within the fixed time slice, and the botnet membership can change withing the time interval. Hence, a multinomial distribution $P(c|a, C_Y)$ has to be estimated for each address $a$ on the training data, based on the model $C_Y$. Note that at application time, $P(c|a, C_Y)$ and hence the right hand side of Equation 2 can only be determined for addresses $a$ that occur in the training data on which $C_Y$ has been inferred.

## 3. Minimal Graph Clustering

Let $X$ be the adjacency matrix of the *input graph* with $n$ nodes. Entries of $X_{ij} = 1$ indicate an edge between nodes $i$ and $j$ which constitutes uncertain evidence for joint membership of these nodes in a botnet. The input matrix is assumed to be reflexive ($X_{ii} = 1$ for all $i$), and symmetric ($X_{ij} = X_{ji}$).

The *outcome* of the clustering process is represented by a reflexive, symmetric *edge selector matrix* $Y$ in which entries of $Y_{ij} = 1$ indicate that nodes $i$ and $j$ are assigned to the same cluster, which indicates that the messages originate from the same botnet. The transitive closure $Y^+$ of matrix $Y$ defines a clustering $C_Y$ of the nodes according to Equation 1. Intuitively, the input matrix $X$ can be thought of as data, whereas output matrix $Y$ should be thought of as the model that encodes a clustering of the nodes. A trivial baseline would be to use $X$ itself as edge selector matrix $Y$. In our application, this would typically lead to all messages being grouped in one single cluster.

No prior knowledge is available on associations between botnets in the absence of empirical evidence. If the adjacency matrix $X$ does not contain evidence that links nodes $i$ and $j$, there is no justification for grouping them into the same cluster. This is reflected in the concept of a *minimal edge selector matrix.*

**Definition 1.** *A selector matrix $Y$ and, equivalently, the corresponding graph clustering $C_Y$, is* minimal with respect to adjacency matrix $X$ *if it satisfies*

$$Y = Y^+ \circ X, \tag{3}$$

*where $(Y^+ \circ X)_{ij} = Y^+_{ij} X_{ij}$ is the Hadamard product that gives the intersection of the edges of $Y^+$ and $X$.*

Intuitively, for every pair of nodes that are connected by the adjacency matrix, selector matrix $Y$ decides whether they are assigned into the same cluster. Nodes that are not connected by the adjacency matrix $X$ must not be linked by $Y$, but can still end up in the same cluster if they are connected by the transitive closure $Y^+$. Equation 3 also ensures that the transitive closure $Y^+$ does not differ from $Y$ for any pair of nodes $i, j$ that are connected by the adjacency matrix. This enforces that no two different minimal selector matrices have the same transitive closures and therefore induce identical clusterings, which would inflate the search space.

### 3.1. Probabilistic Model

This section derives a probabilistic model for the minimal graph clustering problem. Its most salient property is that it is not based on a generative model of the graph, but instead directly models the conditional probability of the clustering given the adjacency matrix $X$. This circumnavigates systematic errors caused by inaccurate distributional assumptions for the generation of the adjacency matrix $X$.

We define the posterior distribution over all reflexive and symmetric matrices $Y$ that are minimal with respect to $X$.

**Definition 2.** *Let $X \in \{0,1\}^{n\times n}$ be a reflexive and symmetric adjacency matrix. Then, $\mathcal{Y}_X \subseteq \{0,1\}^{n\times n}$ is the set of matrices that are reflexive, symmetric, and minimal with respect to $X$.*

In our application, each node is an element of at most two cliques because each message is connected to all other messages that have been sent from the same IP address, and to all other messages that match the same campaign template. If a template or an address has been observed only once, either of these cliques may resolve to just the node itself. Let $Q_X$ denote the set of cliques in $X$, and let $C_Y^q$ be the projection of clustering $C_Y$ to the elements of $q \in Q_X$. Within each clique $q \in Q_X$, any clustering $C_Y^q$ is minimal with respect to $X$ because $X_{ij} = 1$ for all $i, j \in q$, and therefore any reflexive, symmetric, and transitive clustering of $q$ is possible. We model the probability distribution over clusterings of each clique $q \in Q_X$ as a Chinese Restaurant process (Pitman & Picard, 2006) with concentration parameter $\alpha_q > 0$:

$$P(C_Y^q|\alpha_q, n_q) = \alpha_q^{|C_Y^q|} \frac{\Gamma(\alpha_q)}{\Gamma(\alpha_q + n_q)} \prod_{c \in C_Y^q} \Gamma(|c|). \tag{4}$$

Equation 5 now defines the distribution over all partition matrices $Y \in \mathcal{Y}_X$ as a product over all cliques in $Q_X$, where the clique specific concentration parameters are collected into $\boldsymbol{\alpha} = \{\alpha_q : q \in Q_X\}$.

$$P(C_Y|X, \boldsymbol{\alpha}) \propto \begin{cases} \prod_{q\in Q_X} P(C_Y^q|\alpha_q, n_q) & \text{if } Y \in \mathcal{Y}_X \\ 0 & \text{otherwise} \end{cases} \tag{5}$$

Equation 5 can be seen in analogy to the factorization of the posterior over cliques in conditional random fields. However, because the minimality property has non-local effects on the possible values that edges can assume, this factorization is not equivalent to the assumption of the Markov property on which the factorization theorem for random fields is based (Hammersley & Clifford, 1971).

Normalization of Equation 5 is computationally intractable because it requires the enumeration of the elements of $\mathcal{Y}_X$. However, the Gibbs sampler that we will derive in the following only has to normalize over all values of the random variables that are reassigned in each step of the sampling process.

### 3.2. Inference

Computing the posterior distribution $P(C_Y|X, \boldsymbol{\alpha})$ is a generalization of the inference problem for conventional Chinese Restaurant process models. When all entries of $X$ are one, the graph has only one clique and the special case of a Chinese Restaurant process is obtained. In this case, depending on the concentration parameter, the outcome may be one single cluster of all nodes. Maximization of the posterior as well as full Bayesian inference are intractable even for this special case because of the non-convexity of the posterior and the exponential number of possible clusterings. Hence, in this section we describe a Gibbs sampler that generates unbiased samples from the posterior.

**Algorithm 1** Assignment space $\mathcal{Y}_i^Y$ for Gibbs sampler

**Input:** Current partitioning matrix $Y$

let $q_1, \ldots, q_k$ be the cliques with element $i$
let $\mathcal{Y}_i = \emptyset$
**for all** combinations $c_1 \in C_Y^{q_1} \cup \{\{i\}\}$, $\ldots$, $c_k \in C_Y^{q_k} \cup \{\{i\}\}$ **do**
    let $Y'_{-i} = Y_{-i}$
    let $Y'_{il} = Y'_{li} = 1$ if and only if $l \in c_j$ for any $j$
    **if** $(Y'_{-i})^+ = Y'_{-i}$ **then**
        add $Y'$ to $\mathcal{Y}_i^Y$
    **else**
        discard $Y'$
    **end if**
**end for**

**Return:** $\mathcal{Y}_i^Y$, all reflexive, symmetric, minimal partitioning matrices derived from $Y$ by reassigning $Y_i$.

Gibbs samplers divide the set of random variables into smaller subsets and iteratively draw new values for one subset given the values of the remaining variables. For the observations to form an unbiased sample, the random variables have to be partitioned such that the sequence of assignments formes an ergodic Markov chain; that is, each state has to be reachable from each other state. In our case, perhaps the most obvious-seeming approach would be to factor the posterior over individual edges. However, since many matrices $Y$ violate the minimality condition, the chain of alterations of single matrix entries would not in general be ergodic.

Therefore, we devise a sampling algorithm that jointly samples the $i$-th row and column (the $i$-th row and column are identical because $Y$ is symmetric). Let $Y'_i$ refer to the $i$-th row and column of the new matrix $Y'$, and let $Y'_{-i} = Y_{-i}$ refer to the remaining matrix entries, such that $Y' = Y'_i \cup Y'_{-i}$. Equation 6 expands the definition of the conditional probability; Equation 7 factorizes over the cliques, according to Equation 5. Equation 8 omits all terms that are constant in $Y'_i$: the denominator, and all cliques in which node $i$ does not occur. Normalization of the right hand side of Equation 8 is now over all values for $Y'_i$ that render $Y'$ reflexive, symmetric, and minimal with respect to $X$.

$$P(Y'_i|Y_{-i}, X, \boldsymbol{\alpha}) = \frac{P(Y'_i, Y'_{-i}|X, \boldsymbol{\alpha})}{P(Y_{-i}|X, \boldsymbol{\alpha})} \quad (6)$$

$$\propto \frac{\prod_{q \in Q_X} P(C_{Y'}^q|\alpha_q, n_q)}{P(Y_{-i}|X, \boldsymbol{\alpha})} \quad (7)$$

$$\propto \prod_{q \in Q_X : i \in q} P(C_{Y'}^q|\alpha_q, n_q) \quad (8)$$

The main computational challenge here is to determine the set $\mathcal{Y}_i^Y$ of reflexive, symmetric, minimal matrices that can be derived from $Y$ by changing row and column $i$. Since Equation 8 has to be normalized, all of its elements have to be enumerated. An obvious but inefficient strategy would be to enumerate all up to $2^n$ assignments of $Y_i$ and test the resulting matrix for reflexivity, symmetry, and minimality.

However, most values of $Y'_i$ violate minimality and need not be enumerated. Algorithm 1 constructs the set $\mathcal{Y}_i^Y$ in $O(n^k)$, where $k$ is the maximal number of cliques that each node is a member of. In our application, each node is an element of up to two cliques—the set of messages with a shared IP address, and the set of messages that follow the same campaign template. Hence, in our case, the algorithm has a worst-case execution time of $O(n^2)$. In most cases, the number of clusters in each of the two cliques is much lower than $n$, and thus much fewer than $n^2$ cases are considered.

**Theorem 1.** *Given an adjacency matrix $X$ and an edge selector matrix $Y$, Algorithm 1 constructs a set $\mathcal{Y}_i^Y$ that contains all $Y' = Y'_i \cup Y_{-i}$ which are reflexive, symmetric, and minimal with respect to $X$. The execution time of Algorithm 1 is in $O(n^k)$ when each node is a member of at most $k$ cliques in $X$.*

*Proof.* Let node $i$ be an element of cliques $q_1, \ldots, q_k$. On these cliques, the current partitioning matrix $Y$ induces clusterings $C_Y^{q_1}, \ldots, C_Y^{q_k}$ with at most $n$ clusters each. When $Y'_i$ links node $i$ to more than one cluster from any $C_Y^{q_l}$, then by the definition of a clustering in Equation 1 these clusters are merged in $C_{Y'}$. However, when $Y'_i$ links node $i$ to two clusters with at least one other element in $q_l$ each, say $j$ and $l$ with $Y_{jl} = 0$, the transitive closure $Y'^+$ has to add at least an edge to $Y'_{-i}$ that links $j$ and $l$. Since $j$ and $l$ are in clique $q_l$, they have to be connected by the adjacency matrix, $X_{jl} = 1$. But $Y'_{jl} = 0$, $Y'^+_{jl} = 1$ and $X_{jl} = 1$ violates the minimality condition defined in Equation 3. Therefore, $Y'$ must only merge clusters that have elements in different cliques, and so at most $n^k$ combinations of clusters can lead to minimal matrices $Y'$ when merged. Reflexivity and symmetry of $Y'$ follow from reflexivity and symmetry of $X$. The execution time is dominated by the enumeration of all $n^k$ many combinations of clusters in Line 3. □

The Gibbs sampler iteratively samples $Y^{t+1}$ according to $P(Y'_{i_t}|Y^t_{-i_t}, X, \boldsymbol{\alpha})$, given by Equation 8. Each $Y^{t+1}$ is created from the predecessor by cycling over the rows that are resampled—that is, $i_t = t \mod n$. The conditional is defined over the set $\mathcal{Y}_{i_t}^{Y^t}$. We will now argue that a sequence of matrices created by the Gibbs sampler is an ergodic Markov chain.

**Theorem 2.** *For $\alpha_q > 0$, the sequence $Y^0, \ldots, Y^T$ with $Y^{t+1} \sim P(Y^{t+1}_{(t \mod n)}|Y^t_{-(t \mod n)}, X, \boldsymbol{\alpha})$ is an ergodic Markov chain.*

*Proof.* The sequence is a Markov chain because each element is sampled from a distribution that is parameterized only with the preceding matrix and the row that is to be resampled. For it to be ergodic we have to prove that from any state $Y$, every other state $Y'$ can be reached. With the $\alpha_q > 0$, Equation 5 is positive for all states in $\mathcal{Y}_X$ that are reflexive, symmetric, and minimal with respect to $X$. In each step the sampler can only change row and column $i$. Hence, any chain of states with $Y^{t+1} \in \mathcal{Y}^{Y^t}_{(t \mod n)}$ can be reached because by Theorem 1, all elements of $\mathcal{Y}^{Y^t}_{(t \mod n)}$ are reflexive, symmetric, minimal with respect to $X$ and differ from $Y^t$ only in row and column $i$.

To begin with, we argue that from any state $Y$ the identity matrix $I$ can be reached which connects each node only to itself. To prove this, it suffices to show that for any $i$ and any $Y$, a state $I^{Y,i}$ with $I^{Y,i}{}_{ii} = 1$ for all $i$, $I^{Y,i}{}_{ij} = 0$ for all $j \neq i$, and $I^{Y,i}{}_{jk} = Y_{jk}$ for all $j, k \neq i$ can be reached directly from state $Y$ by sampling row and column $Y_i$. By the definition of $\mathcal{Y}_i^Y$, the Gibbs sampler can directly reach state $I^{Y,i}$ from $Y$ if $I^{Y,i}$ is symmetric, reflexive, minimal with respect to $X$, and differs from $Y$ only in the $i$-th row and column. By its definition, it is clear that $I^{Y,i}$ differs from $Y$ only in the $i$-th column and row, and that it is reflexive. Since $Y$ is symmetric and the $i$-th row and column of $I^{Y,i}$ are identical, $I^{Y,i}$ has to be symmetric as well. It remains to be shown that $I^{Y,i} = I^{Y,i+} \circ X$. We split the proof of this claim into two parts. First, we show that the $i$-th row and column of $I^{Y,i}$ are equal to the $i$-th row and column of $I^{Y,i+} \circ X$. Intuitively, because $I^{Y,i}$ connects node $i$ only to itself, the transitive closure adds nothing, and the Hadamard product has no effect because $X$ is reflexive. Formally, this can be shown via an inductive proof along the following construction of the transitive closure of $I^{Y,i}$. Let $R^0 = I^{Y,i}$. For all $l > 0$, let $R^l_{ij} = 1$ if $R^{l-1}_{ij} = 1$ or if there is a $k$ such that $R^{l-1}_{ik} = 1$ and $R^{l-1}_{kj} = 1$; otherwise, $R^l_{ij} = 0$. When $R^{l-1}$ contains an open triangle of edges $R^{l-1}_{ik} = 1$ and $R^{l-1}_{kj} = 1$, then $R^l$ is defined to add an edge $R^l_{ij} = 1$. Then the limit $\lim_{l \to \infty} R^l$ is the transitive closure $(I^{Y,i})^+$. Now inductively, if for all $j : R^{l-1}_{ij} = 0$, then for all $j : R^l_{ij} = 0$, and from $I^{Y,i}{}_i = 0$ it follows that $I^{Y,i+}_i = I^{Y,i}{}_i$, and $(I^{Y,i+} \circ X)_i = I^{Y,i}{}_i$.

Secondly, we show that all elements in $I^{Y,i+}$ *except* the $i$-th row and column remain unchanged from $Y^+$: From the monotonicity of the transitive closure operator and $I^{Y,i} \leq Y$ it follows that $(I^{Y,i+} \circ X)_{-i} \leq (Y^+ \circ X)_{-i}$. Furthermore, since the transitive closure operator only adds positive edges, $(I^{Y,i+} \circ X)_{-i} \geq (I^{Y,i} \circ X)_{-i} = (Y \circ X)_{-i}$, which is in turn equal to $(Y^+ \circ X)_{-i}$ because $Y$ itself is minimal with respect to $X$. Both inequalities together give us $(I^{Y,i+} \circ X)_{-i} = (Y^+ \circ X)_{-i}$, and because $I^{Y,i}{}_{-i} = Y_{-i}$ we have that $I^{Y,i}{}_{-i} = (I^{Y,i+} \circ X)_{-i}$. Together with the first part finally $I^{Y,i} = I^{Y,i+} \circ X$.

This establishes that $I^{Y,i}$ can be reached by the Gibbs sampler from any state $Y$ for any $i$, and thus by repeatedly using this state transition for *all* $i$, $I$ is reachable. The reachability relation is symmetric because $Y^{t+1}$ is constructed from $Y^t$ by reassigning one column and row which can be reversed, and $Y^t$ is required to be in $\mathcal{Y}_X$, and therefore can be reached from $Y^{t+1}$. Hence, from any state $Y$, every other state $Y'$ can be reached via the state $I$, and ergodicity holds. □

### 3.3. Prediction

The Gibbs sampler creates a chain $Y^0, \ldots, Y^T$ of matrices, governed by the posterior $P(Y|X, \boldsymbol{\alpha})$. In order to predict which campaign $s$ a given IP address $a$ will participate in, we can approximate the Bayesian infer-

ence of $c$ (Equation 9) using the chain (Equation 10).

$$P(s|a, X, \boldsymbol{\alpha}) = \sum_{Y \in \mathcal{Y}_X} P(C_Y|X, \boldsymbol{\alpha}) P(s|a, C_Y) \quad (9)$$

$$\approx \sum_{Y \in \{Y^0, \ldots, Y^T\}} P(s|a, C_Y) \quad (10)$$

Equation 1 decomposes $P(s|a, C_Y)$ into two multinomial distributions that can be estimated from the available data.

## 4. Case Study

In this section, we conduct a case study on botnet detection. Since the gound truth about which botnets are currently active and which hosts they are composed of is not available, we evaluate the model in terms of its accuracy of predicting which spam campaign a given IP address will participate in.

We record incoming spam emails over a period of 11 days in January 2012 at a large email service provider. We select only emails that have been blacklisted on the grounds of three content-based filtering techniques: The first is a set of manually maintained regular expressions, each tailored to match against all spams of one particular campaign. The second is a list of semi-automatically generated, campaign-specific feature sets (Haider & Scheffer, 2009). A feature set consists of words and structure flags and is the intersection of all previously observed emails from the campaign. The third is a blacklist of URLs that spam emails link to. Thus, we have a reliable partitioning of all emails into spam campaigns.

We exclude IP addresses of known legitimate forwarding servers that relay inbound emails according to their users' personal filtering policies. To this end, we track the IP address from the last hop in the transmission chain. If the address has a valid reverse DNS entry that matches a domain from a list of well-known email service providers, we omit the message.

Equation 5 allows for individual values of the concentration parameters $\alpha_q$ for each clique $q$ in the email graph $X$. We use two distinct values: a value of $\alpha_a$ for all cliques that share a joint IP address, and a value of $\alpha_s$ for all cliques that match a joint campaign. Parameters $\alpha_a$ and $\alpha_s$ are tuned to maximize the AUC metric on the data recorded on the first day. The data of the remaining ten days is then used for evaluation. Within each day, the Gibbs sampler infers a chain of clusterings on the data of the first 16 hours. The emails of the last 8 hours with a sender IP address that has previously occurred are used as test data. Emails from IP addresses that have not been seen before are excluded, since no informed decision can be made for them. The proportion of IP addresses that have not previously been observed depends on the proportion of the global email traffic that the server gets to observe. Also, we exclude emails from campaigns that appear less than 100 times. In total, this data collection procedure results in 701,207 unique pairs of campaigns and IP addresses in the training sets and 71,528 in the test sets. Each test email serves as a positive example for its campaign and a negative example for all other campaigns.

### 4.1. Reference Methods

We compare the Minimal Graph Clustering model to three baselines. The first, *threshold-based* baseline is an agglomerative clustering algorithm based on a threshold heuristic, adapted from Zhuang et al. (2008). It operates on the assumption that each campaign is sent by only one botnet. Initially, every campaign constitutes its own cluster. Clusters $c$ and $c'$ are greedily merged if their fraction of overlapping IP addresses exceeds a threshold. This fraction is defined as

$$\frac{\sum_{i \in c} \mathbb{I}(\exists j \in c' : s_i = s_j)}{2|c|} + \frac{\sum_{j \in c'} \mathbb{I}(\exists i \in c : s_j = s_i)}{2|c'|},$$

where $\mathbb{I}$ is the indicator function and $s_i$ the campaign of the $i$-th email. Given a clustering $C$ of emails, $P(s|a, C) = \sum_{c \in C} P(s|c, C)P(c|a, C)$ is inferred after multinomial distributions $P(s|c)$ and $P(c|a, C)$ have been estimated on the training data. The clustering threshold is tuned for performance on the first day.

The second baseline is *spectral clustering*, where we tune the number of clusters and similarity values for emails with matching campaign or IP address. We use the implementation of Chen et al. (2011).

The third baseline is a straightforward generative clustering model for email graphs with a Chinese Restaurant process prior and a likelihood function that factorizes over the edges of the email graph $X$, assuming independence for the edges in $X$. The likelihood function has a set of four parameters $\boldsymbol{\theta} = \{\theta_s^{in}, \theta_s^{in}, \theta_a^{out}, \theta_a^{out}\}$ that quantify the probability of the presence of a link when the nodes are and are not elements of a joint botnet. The likelihood for an edge that connects two emails from the same campaign is given as

$$P(X_{ij} = 1|C, \boldsymbol{\theta}) = \begin{cases} \theta_s^{in}, & \text{if } C(i) = C(j) \\ \theta_s^{out}, & \text{if } C(i) \neq C(j), \end{cases}$$

where $C(i)$ denotes the cluster that clustering $C$ assigns email $i$ to. The likelihood of an edge between two emails from the same IP address is defined analogously, using the paramters $\theta_a^{out}$ and $\theta_a^{out}$. The joint probability of the email matrix and a clustering $C$ is then given as $P(X, C|\alpha, \boldsymbol{\theta}) = P_{CRP}(C|\alpha) \prod_{i,j<i} P(X_{ij}|C, \boldsymbol{\theta})$. The parameters are adjusted using gradient-ascent on the joint data likelihood.

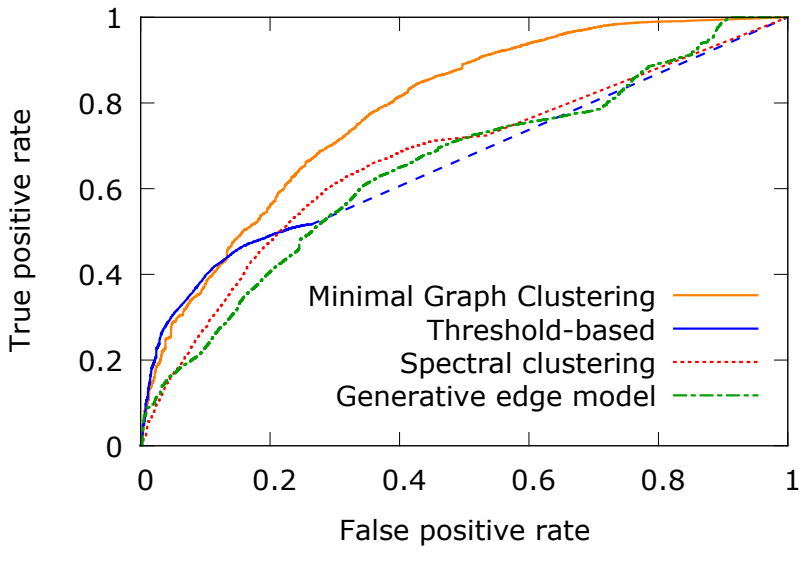

*Figure 1.* ROC-curves for campaign prediction.

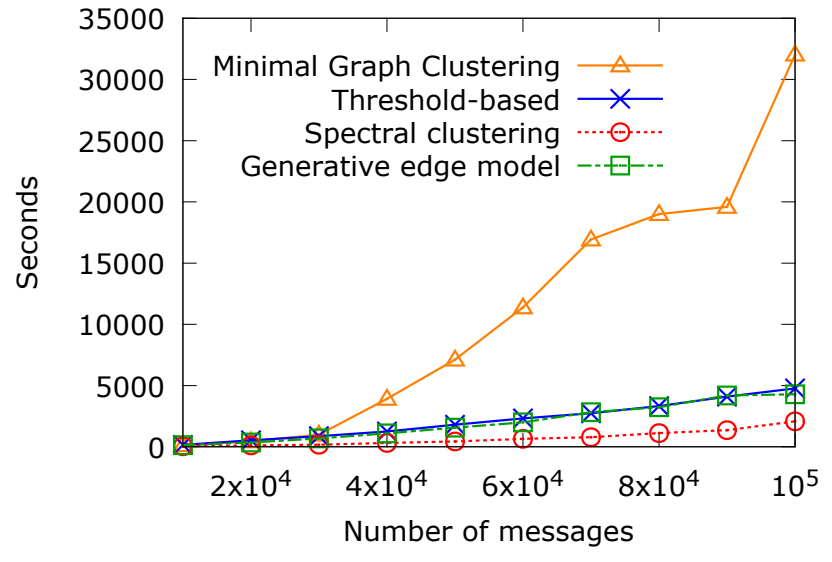

*Figure 2.* Execution times for inferring the clustering.

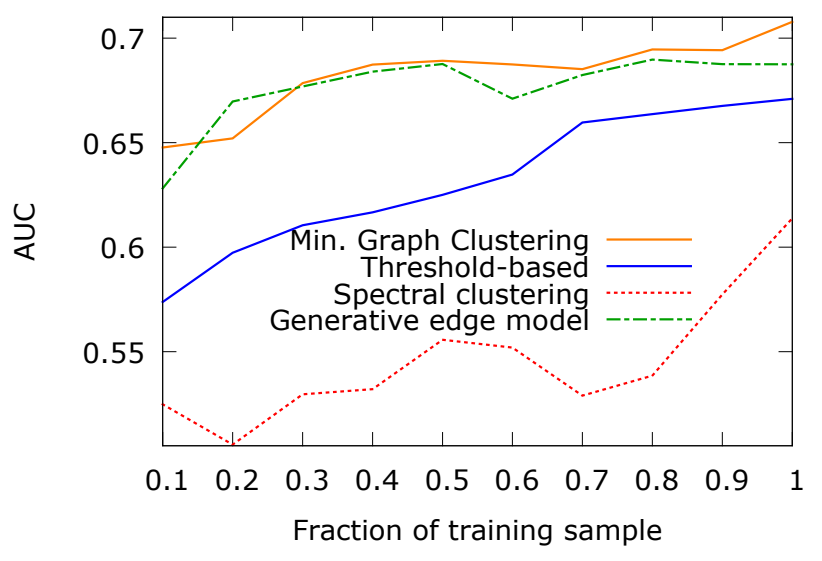

*Figure 3.* AUC depending on training set size.

### 4.2. Results

We measure ROC curves; IP addresses for which the likelihood $P(s|a, X, \boldsymbol{\alpha})$ of the correct campaign exceeds the threshold are counted as true positives, addresses for which the likelihood of an incorrect campaign exceeds the threshold as false positives. Figure 4 shows ROC curves for the four methods under study.

Minimal graph clustering attains the highest area under the ROC curve of 0.784, compared to 0.675 for threshold-based clustering, 0.671 for spectral clustering, and 0.651 for the generative edge model baseline. The threshold-based method is marginally more accurate than the Minimal Graph Clustering model for low threshold values, and less accurate for all other threshold values. Threshold-based clustering infers $P(s|c)$ and $P(c|a, C)$ and consequently $P(s|a, C)$ to be zero for many values of $s$ and $a$. Therefore, a large interval of points on the ROC curve cannot be attained by any threshold value; this is indicated by a dashed line.

Typically, the number of requests during a DDoS attack exceeds the capacity to serve requests by far. An unprotected system will serve only a small, random fraction of requests and will be unable to serve all others; this amounts to a false positive rate of close to one. In order to defend against such an attack, one has to select a small proportion of requests which can be served. Therefore, in defending against DDoS attacks, the right hand side of the ROC curve that allows high true positive rates is practically relevant.

Figure 4 shows execution times for running all four methods until convergence depending on the number of examples. The Minimal Graph Clustering model is computationally more expensive than the baselines. For continuously maintaining a clustering that subsequently incorporates newly available messages, it is thus advisable to use the previous clustering as a starting point of the Gibbs sampler in order to reduce the number of necessary iterations until convergence.

Figure 4 shows area under ROC curve depending on what fraction of the training sample is used. For testing, only emails with IP addresses that are present in the smallest subset are used. The plots indicate that having access to a larger sample of the overall email traffic could increase performance considerably.

## 5. Conclusion and Discussion

The identification of spam-disseminating botnets can be reduced to the problem of clustering the graph of email messages in which messages are linked if they originate from the same IP address or match the same campaign template. We devised a probabilistic model that directly describes the conditional probability of a clustering given the input graph without making distributional assumptions about the generation of the observable data. We derived a Gibbs sampler; we showed that resampling rows and edges of the output matrix creates an ergodic Markov chain, and that each sampling step can be carried out in $O(n^2)$. We argue that botnet models can be evaluated in terms of their ability to predict which spam campaign a given IP address is going to participate in. From a case study carried out with an email service provider we conclude that the minimal graph clustering model outperforms a number of reference methods—spectral clustering, a generative model, and a threshold-based, agglomerative clustering model—in terms of its area under the ROC curve.

The botnet model draws a picture of the current size and activity of botnets. From the IP addresses, the geographical distribution of each botnet can be derived. The botnet model can be used to select particularly prolific botnets for in-depth analysis and possible legal action. Widespread botnet software is versatile and supports both, dissemination of email spam and the staging of network attacks (Stern, 2008). When both, a mailing campaign and a network attack are carried out by a single network within the typical IP-address reassignment interval of one day, then the botnet model which has been trained on email data can score HTTP requests by the likelihood that their sender IP address is part of an attacking botnet. This allows to prioritize requests and to maintain a service during an attack. Alternatively, the botnet model can be trained with HTTP requests instead of emails; the recipient domain of an HTTP request plays the role of the campaign template. Again, the botnet model allows to infer the likelihood that an individual sender IP address acts as part of an attacking botnet.

Direct evaluation of the model's ability to decide whether an IP request is part of a network attack would require evaluation data in the form of a collection of individual HTTP requests labeled with the botnet that has sent the request. While it is relatively easy to collect the entire stream of legitimate and attacking HTTP requests that reach a domain during an attack, there is no practical means of labeling individual requests. In general, HTTP requests contain no information that allows even a human expert to decide whether a request is part of an attack, let alone which botnet a request has really been sent from.

## 6. Acknowledgments

This work was funded by a grant from STRATO AG.